\documentclass[english,12pt]{article}
\usepackage{array}
\usepackage{graphicx}
\usepackage{amssymb}
\usepackage[english]{babel}
\usepackage{amsmath}
\usepackage{multirow}
\usepackage{prettyref}
\usepackage{babel}
\usepackage{units}
\usepackage[latin1]{inputenc}
\usepackage{amsfonts}
\usepackage{amssymb}
\usepackage{babel}
\usepackage{color}

\usepackage{xcolor}
\usepackage{hyperref}
\hypersetup
{colorlinks   = true, 
 urlcolor     = black,
 linkcolor    = black, 
 citecolor    = black}

\usepackage{amsfonts}
\usepackage{amssymb}
\usepackage{amsmath}
\usepackage{amsthm}
\usepackage{url}
\usepackage{comment}
\usepackage{physics}

\usepackage{cite}
\def\@fmsl@sh#1#2#3{\m@th\ooalign{$\hfil#1\mkern#2/\hfil$\crcr$#1#3$}}
 \def\eq#1\en{\begin{equation}#1\end{equation}}
\def\s[#1,#2]{[#1\stackrel{\star}{,}#2]}
\def\sx[#1,#2]{[#1\stackrel{\star_{x}}{,}#2]}

\def\bc{\begin{center}}
\def\ec{\end{center}}

\def\gsim{\mathrel{\mathpalette\atversim>}}

\def\bc{\begin{center}}
\def\ec{\end{center}}

\def\gsim{\mathrel{\rlap{\lower4pt\hbox{\hskip1pt$\sim$}}

    \raise1pt\hbox{$>$}}}       

\def\gsim{\mathrel{\rlap{\lower4pt\hbox{\hskip1pt$\sim$}}
    \raise1pt\hbox{$>$}}}       

\begin{document}
\makeatletter
\def\fmslash{\@ifnextchar[{\fmsl@sh}{\fmsl@sh[0mu]}}
\def\fmsl@sh[#1]#2{%
  \mathchoice
    {\@fmsl@sh\displaystyle{#1}{#2}}%
    {\@fmsl@sh\textstyle{#1}{#2}}%
    {\@fmsl@sh\scriptstyle{#1}{#2}}%
    {\@fmsl@sh\scriptscriptstyle{#1}{#2}}}
\def\@fmsl@sh#1#2#3{\m@th\ooalign{$\hfil#1\mkern#2/\hfil$\crcr$#1#3$}}
\makeatother

\thispagestyle{empty}
\begin{titlepage}
\boldmath
\begin{center}
  \Large {\bf  Quantum Gravitational Corrections to Particle Creation by Black Holes
}
    \end{center}
\unboldmath
\vspace{0.2cm}
\begin{center}
{{\large Xavier~Calmet}\footnote{E-mail: x.calmet@sussex.ac.uk}$^a$,
 {\large  Stephen D. H. Hsu}\footnote{E-mail: hsusteve@gmail.com}$^b$}
{\large and} 
{\large Marco~Sebastianutti}\footnote{E-mail: m.sebastianutti@sussex.ac.uk}$^a$
 \end{center}
\begin{center}
$^a${\sl Department of Physics and Astronomy, \\
University of Sussex, Brighton, BN1 9QH, United Kingdom
}\\
$^b${\sl Department of Physics and Astronomy\\ Michigan State University, East Lansing, Michigan 48823, USA}\\
\end{center}
\vspace{2cm}
\begin{abstract}
\noindent
We calculate quantum gravitational corrections to the amplitude for the emission of a Hawking particle by a black hole. We show explicitly how the amplitudes depend on quantum corrections to the exterior metric (quantum hair). This reveals the mechanism by which information escapes the black hole. The quantum state of the black hole is reflected in the quantum state of the exterior metric, which in turn influences the emission of Hawking quanta.

\end{abstract}
\vspace{5cm}
\end{titlepage}

\section{Introduction}

In this paper we compute quantum gravitational corrections to the amplitudes for creation of particles by black holes, first discussed by Hawking \cite{Hawking:1975vcx}. 

We show that the Hawking amplitudes are modified by quantum corrections to the exterior metric (i.e., quantum hair). This reveals, explicitly, a mechanism by which information escapes from black holes. The exterior metric state is itself determined by the quantum state of the black hole, so there is a direct connection between the particle states produced during evaporation and the history of the black hole.

In a series of recent papers \cite{Calmet:2017qqa,Calmet:2018elv,Calmet:2019eof,Calmet:2021lny,Calmet:2021stu,Calmet:2021cip,Calmet:2022swf,Calmet:2022bpo}, it has been shown that quantum gravitational corrections generically lead to quantum hair. In other words, classical solutions to Einstein's equations generically receive quantum gravitational corrections. As long as one considers physical processes taking place at  small curvature, which is the case for all practical purposes in our universe except very close to gravitational singularities, these corrections can be calculated in a reliable manner using the unique effective action approach \cite{Barvinsky:1983vpp,Barvinsky:1985an,Barvinsky:1987uw,Barvinsky:1990up,Buchbinder:1992rb}. 

We will first revisit Hawking's calculation \cite{Hawking:1975vcx} and then consider the tunneling method of Parikh and Wilczek \cite{Parikh:1999mf}. The latter accounts for energy conservation -- i.e., that the black hole mass and radius shrink in the evaporation process -- whereas Hawking's methods assume a purely static black hole. We first consider Schwarzschild black holes before extending our results to a generic collapse model.

\section{Quantum Gravitational corrections to Hawking radiation}

It has been shown in \cite{Calmet:2017qqa,Calmet:2018elv,Calmet:2021lny} that the leading order quantum gravitational correction to the Schwarzschild metric appears at third order in curvature.  It is generated by a local operator $c_6  G_N R^{\mu\nu}_{\ \ \alpha \sigma} R^{\alpha\sigma}_{\ \ \delta \gamma} R^{\delta \gamma}_{\ \ \mu\nu}$, where $G_N$ is Newton constant. As this is a local operator, its Wilson coefficient is not calculable from first principles in the approach, but it must be extracted from experiment or an observation. The leading order quantum gravitational correction to Schwarzschild metric is \cite{Calmet:2021lny}
\begin{align}
    ds^2&=-f(r)dt^2+\frac{1}{g(r)}dr^2+r^2d\Omega^2\,,\label{eq:sph}\\
    f(r)&=1-\frac{2G_N M}{r}+640\pi c_6 \frac{G_N^5 M^3}{r^7}\label{eq:fqs}\,,\\
    g(r)&=1-\frac{2G_N M}{r}+128\pi c_6 \frac{G_N^4 M^2}{r^6}\Big(27-49\frac{G_N M}{r}\Big)\,.\label{eq:gqs}
\end{align}
We will now show how this quantum correction impacts the emission of Hawking particles.

In his seminal paper \cite{Hawking:1975vcx}, Hawking considers the wave-function of a scalar field  $\Box \Phi=0$  in a curved space-time given by the Schwarzschild metric.
The field operator can be written as
\begin{eqnarray}
\Phi=\sum_i \left (f_i {\bf a_i}+\bar f_i {\bf a^\dagger_i} \right)=\sum_i \left (p_i {\bf b_i} + \bar p_i {\bf b^\dagger_i} + q_i {\bf c_i} + \bar q_i {\bf c^\dagger_i} \right ) \, ,
\end{eqnarray}
where $f_i$, $\bar f_i$ (the bar denotes complex conjugation) are solutions of the wave equation which are purely ingoing, whereas $p_i$, $\bar p_i$ and $q_i$, $\bar q_i$ are respectively purely outgoing solutions and solutions which do not contain any outgoing component. The ${\bf a_i}$, ${\bf b_i}$ and ${\bf c_i}$ are annihilation operators and ${\bf a^\dagger_i}$, ${\bf b^\dagger_i}$ and ${\bf c^\dagger_i}$ are creation operators.

Our aim is to show that the solutions $f_i$, $\bar f_i$, $p_i$, $\bar p_i$, $q_i$ and $\bar q_i$ depend on the quantum corrections.  Indeed, we will show that quantum gravitational corrections to the Schwarzschild metric lead to modifications of Hawking's solutions. This implies that Hawking radiation carries information about the quantum system which is the source of the quantum hair. 

The classical Schwarzschild space-time and its quantum corrections given above are spherically symmetric. One can thus decompose the incoming and outgoing solutions into spherical harmonics. In the region outside the black hole, one can write the incoming and outgoing solutions as
\begin{eqnarray}
f_{\omega^\prime l m} &=& \frac{1}{\sqrt{2 \pi \omega^\prime} r }  F_{\omega^\prime}(r) e^{i \omega^\prime v} Y_{lm}(\theta,\phi)\ , \\ 
p_{\omega l m} &=& \frac{1}{\sqrt{2 \pi \omega} r }  P_\omega(r) e^{i \omega u} Y_{lm}(\theta,\phi) \, 
\end{eqnarray}
where $v$ and $u$ are the advanced and retarded coordinates. In the  classical  Schwarzschild case, they are given by
\begin{eqnarray}
v^c&=&t+r + 2 M \log \left \lvert \frac{r}{2 M G_N} -1 \right \rvert \\ 
u^c&=&t-r - 2 M \log \left \lvert \frac{r}{2 M G_N} -1 \right \rvert.
\end{eqnarray}
Our aim is to find the leading quantum correction to these functions. The simplest way to do so is to consider a geodesic path for a particle moving in the background space-time, parametrized by an affine parameter $\lambda$. The 4-momentum of the particle is given by 
\begin{eqnarray}\label{eq:4mom}
    p_\mu=g_{\mu\nu}\frac{dx}{d\lambda}^\nu
\end{eqnarray}
and is conserved along geodesics. Furthermore, the quantity
\begin{eqnarray}\label{eq:cons}
\epsilon=-g_{\mu\nu}\frac{dx}{d\lambda}^\mu \frac{dx}{d\lambda}^\nu \,,
\end{eqnarray}
is also conserved along geodesics. For massive particles we set $\epsilon = 1$ and $\lambda=\tau$ is the proper time along the path. For massless particles, which is the case we are going to examine, $\epsilon = 0$ and $\lambda$ is an arbitrary affine parameter.

Considering a generic stationary spherically symmetric metric as in~\eqref{eq:sph}, and taking radial ($p_\phi=L=0$) geodesics on the plane specified by $\theta=\pi/2$, from~\eqref{eq:4mom} we get
\begin{eqnarray}
    E&=& f(r) \dot{t} 
\end{eqnarray}
where $E=-p_t$ and a dot stands for $d/d\lambda$. From~\eqref{eq:cons} we instead obtain
\begin{eqnarray}\label{eq:drdl}
    \left( \frac{dr}{d\lambda} \right)^2 &=& \frac{E^2}{f(r)g(r)^{-1}}.
\end{eqnarray}
Manipulating the two equations above we arrive at 
\begin{eqnarray}\label{eq:uv}
    \frac{d}{d\lambda}\left(t\mp r^*\right) &=& 0
\end{eqnarray}
where $r^*$ is the tortoise coordinate defined as
\begin{eqnarray}
    dr^* &=& \frac{dr}{\sqrt{f(r)g(r)}}\,.
\end{eqnarray}
The two conserved quantities in~\eqref{eq:uv} are the advanced and retarded coordinates, $v$ and $u$ respectively. Rewriting the expression for this latter coordinate we finally obtain
\begin{eqnarray}\label{eq:dudl}
    \frac{du}{d\lambda}=\frac{2E}{f(r)}\,.
\end{eqnarray}
Along an ingoing geodesic $\mathcal{C}$ parametrized by $\lambda$, the advanced coordinate $u$ is expressed as some function $u(\lambda)$. Only two steps are needed to get an expression for such a function: we first write $r$ as a function of $\lambda$ and then perform the integral in \eqref{eq:dudl}.
The form of $u(\lambda)$ determines the final expressions of the Bogoliubov coefficients, which are ultimately responsible for the emission of quanta by the black hole.

We now take $f(r)$ and $g(r)$ as in~\eqref{eq:fqs} and~\eqref{eq:gqs} and then perform the integral of the square root of~\eqref{eq:drdl} in the interval $r'\in\left[r_H,r\right]$ corresponding to $\lambda'\in\left[0,\lambda\right]$.\footnote{Note that in the square root of~\eqref{eq:drdl} we pick the minus sign corresponding to the ingoing geodesic $\mathcal{C}$.} Assuming that quantum corrections are small compared to the classical Schwarzschild terms, and remaining close enough to the horizon, we find
\begin{eqnarray}\label{eq:rl}
    r=r_H-E\lambda\left(1+\frac{1728\pi G_N^4M^2}{r_H^6}c_6\right)+{\cal O}\left(\lambda^2\right)+{\cal O}\left(c_6^2\right),
\end{eqnarray}
where $r_H$ is the modified Schwarzschild radius, defined by $g(r_H)=0$:
\begin{eqnarray}
    r_H=2G_N M\left(1-\frac{5\pi}{G_N^2 M^4}c_6\right)+ {\cal O}\left(c_6^2\right)\,.
\end{eqnarray}
We can now use $r(\lambda)$ in~\eqref{eq:rl} to integrate~\eqref{eq:dudl}, hence obtaining
\begin{equation}\label{eq:ul}
    u(\lambda)=-4 M G_N\log (\frac{\lambda}{C} )+\frac{8 \pi  c_6}{M^3 G_N}\log (\frac{\lambda}{C})+{\cal O}\left(\lambda\right)+{\cal O}\left(c_6^2\right),
\end{equation}
with $C$ a negative integration constant. Geometric optics then allows us to relate the outgoing null coordinate to the ingoing one: $\lambda=(v_0-v)/D$, where $v_0$ is the advanced coordinate which gets reflected into the horizon ($\lambda=0$) and $D$ is a negative constant.

We can now calculate the quantum corrected out-going solutions to the original Klein-Gordon equation. They are given by
\begin{eqnarray}
p_{\omega} =\int_0^\infty \left ( \alpha_{\omega\omega^\prime} f_{\omega^\prime} + \beta_{\omega\omega^\prime} \bar f_{\omega^\prime}  \right)d\omega^\prime,
\end{eqnarray}
where $\alpha_{\omega\omega^\prime}$ and $\beta_{\omega\omega^\prime}$ are the
 Bogoliubov coefficients given by
\begin{multline}
    \alpha_{\omega\omega^\prime}=-iKe^{i\omega^\prime v_0}e^{\left(2\pi MG_N-\frac{4\pi^2c_6}{M^3G_N}\right)\omega}\\
    \times \int_{-\infty}^{0} \,dx\,\Big(\frac{\omega^\prime}{\omega}\Big)^{1/2}e^{\omega^\prime x}\text{exp}\left[i\omega\Big(4MG_N-\frac{8\pi c_6}{M^3G_N}\Big)\text{ln}\left(\frac{|x|}{CD}\right)\right]
\end{multline}
and
\begin{multline}
    \beta_{\omega\omega'}=iKe^{-i\omega^\prime v_0}e^{-\left (2\pi MG_N-\frac{4\pi^2c_6}{M^3G_N}\right)\omega}\\
    \times \int_{-\infty}^{0} \,dx\,\Big(\frac{\omega^\prime}{\omega}\Big)^{1/2}e^{\omega^\prime x}\text{exp}\left[i\omega\Big(4MG_N-\frac{8\pi c_6}{M^3G_N}\Big)\text{ln}\left(\frac{|x|}{CD}\right)\right]\,.
\end{multline}
Note that these integrals can be represented in terms of Gamma functions.

This demonstrates that the quantum amplitude for the production of the particle depends on the quantum correction to the metric. In other words, the quantum hair induces a quantum correction to the amplitude for production of Hawking radiation. This is the mechanism by which information escapes the black hole.

Note that while the quantum amplitude receives a quantum gravitational correction, the power spectrum at this stage is still that of a black body. To show this, it suffices to calculate
\begin{equation}\label{eq:alphabetarel}
    |\alpha_{\omega\omega'}|^2=\text{exp}\Big[\big(8\pi G_NM-\frac{16\pi^2c_6}{M^3G_N}\big)\omega\Big]|\beta_{\omega\omega'}|^2\,.
\end{equation}
From the flux of outgoing particles with frequencies between $\omega$ and $\omega + d\omega$ \cite{Kraus:1994by}:
\begin{equation}\label{Fomega}
    F(\omega)=\frac{d\omega}{2\pi}\frac{1}{\left \lvert\frac{\alpha_{\omega\omega^\prime}}{\beta_{\omega\omega^\prime}}\right \rvert^2-1}\, .
\end{equation}
We finally arrive at:
\begin{equation}\label{PTm}
    F(\omega)=\frac{d\omega}{2\pi}\frac{1}{e^{\big(8\pi G_NM-\frac{16\pi^2c_6}{M^3G_N}\big)\omega}-1}\,.
\end{equation}
Comparing with the Planck distribution
\begin{equation} \label{N}
    F(\omega)=\frac{d\omega}{2\pi}\frac{1}{e^{\frac{\omega}{T}}-1}
\end{equation}
we find that 
\begin{equation}\label{eq:TH2}
    T=\frac{1}{8\pi M G_N\Big(1-\frac{2\pi c_6}{M^4G_N^2}\Big)}=\frac{1}{8\pi M G_N}\Big[1+2\pi c_6\Big(\frac{1}{G_N^2M^4}\Big)\Big]+{\cal O}(c_6^2)\, ,
\end{equation}
which is the same expression as the expression for the temperature obtained using the surface gravity in~\cite{Calmet:2021lny}. Eq.~\eqref{N} implies that for a quantum corrected Schwarzschild metric, a black hole emits radiation exactly like a gray body of temperature $T$ given by \eqref{eq:TH2}.\vspace{10pt}

However, we have not yet imposed energy conservation of the whole system. When the black hole radiates a particle its total mass decreases and the black hole shrinks. In the next section we follow the tunneling approach of Wilczek and Parikh \cite{Parikh:1999mf}, which takes this effect into account.

\section{Quantum Gravitational corrections to the tunneling process}
\label{sec:3}
We now follow \cite{Parikh:1999mf,Vanzo:2011wq,Parikh:2004rh} to implement energy conservation in the derivation of the spectrum of radiation for our quantum corrected Schwarzschild metric.  The quantum corrected Schwarzschild metric in Painlev\'e-Gullstrand form  is easily derived:
\begin{eqnarray}
    ds^2&=&-f(r)dt^2+2h(r)dtdr+dr^2+r^2d\Omega^2\,,
    \end{eqnarray}
    where
    \begin{eqnarray}
    h(r)&=& \sqrt{f(r)\big(g(r)^{-1}-1\big)}\\  &=& \nonumber
    \sqrt{\frac{2G_N M}{r}} \left(1-\left(864 \pi\frac{G_N^3 M }{r^5}+160 \pi\frac{G_N^4  M^2 }{r^6}\right)c_6+{\cal O}(c_6^2)\right)\,.
\end{eqnarray}

The tunneling rate is related to the imaginary part of the action \cite{Parikh:1999mf,Vanzo:2011wq,Parikh:2004rh}. 
We can write the action for a particle moving freely in a curved background as
\begin{equation}\label{action}
    S=\int \, p_\mu dx^\mu
\end{equation}
with $p_\mu$ as in~\eqref{eq:4mom}. As we are computing Im$\,S$, the second term in $p_\mu dx^\mu=p_tdt+p_rdr$ does not contribute, since $p_t\,dt=-E dt$ is real, hence
\begin{equation}\label{eq:Ims}
    \text{Im}\,S=\text{Im}\,\int_{r_i}^{r_f} \,p_r\,\text{d}r=\text{Im}\,\int_{r_i}^{r_f}\int_{0}^{p_r} \,\text{d}p_r'\,\text{d}r\,.
\end{equation}
Using Hamilton's equation and  the Hamiltonian of the system which is given by $H=M-E^\prime$, we find $dH=-(dE^\prime)$, with $0\leq E^\prime \leq E$ where $E$ is the energy of the quantum emitted. We thus obtain:
\begin{eqnarray}
\text{Im}\,S=\text{Im}\,\int_{r_i}^{r_f}\int_{M}^{M-E} \,\frac{dH}{dr/dt}\,dr
    =\text{Im}\,\int_{r_i}^{r_f}\,dr\int_{0}^{E} \,\frac{(-dE')}{dr/dt}\,.
\end{eqnarray}
Switching the order of integration and substituting, 
\begin{eqnarray}
    \frac{\text{d}r}{\text{d}t}&=&-h(r)+\sqrt{f(r)+h(r)^2}~=~1-\sqrt{\frac{\Delta(r)}{r}}\, ,
\end{eqnarray}
where 
$
\Delta(r)=(\sqrt{2G_NM}+\sqrt{r}\delta(r))^2$ with the quantum correction given by
\begin{eqnarray}
    \delta(r)&=&32\pi\frac{G_N^3M}{r^6}\left(-54MG_N+\sqrt{\frac{2MG_N}{r}}\left(27r+5MG_N\right)\right)c_6+\mathcal{O}(c_6^2)\, ,
\end{eqnarray}
we find
\begin{eqnarray}
    \text{Im}\,S=\text{Im}\,\int_{0}^{E} \,(-dE')\int_{r_i}^{r_f}\,\frac{dr}{1-\sqrt{\frac{\Delta(r,\,E^\prime)}{r}}}\,.
\end{eqnarray}
The function $\Delta(r)$ is now a function of $E^\prime$ following the substitution $M\rightarrow (M-E^\prime)$ in the original metric. This integral has a pole at the new horizon $r=r_H$, which is obtained by solving $g(r_H)=0$. Integrating along a counterclockwise contour we finally obtain:
\begin{eqnarray}
    \text{Im}\,S=-2 \pi  E  G_N (E -2 M)+\frac{8 \pi ^2 c_6}{G_N} \left(\frac{1}{2 M^2}-\frac{1}{2 (E -M)^2}\right)\, .
\end{eqnarray}
Following the derivation in \cite{Vanzo:2011wq}, the quantum corrected emission rate of a Hawking particle is given by
\begin{eqnarray}
    \Gamma \sim e^{-2 \, \text{Im}\, S}=e^{-8 \pi  E M  G_N \left (1-\frac{E}{2 M}\right)
    \left (1-\frac{2\pi c_6}{G_N^2 M^2\left(M-E\right)^2}\right )} \, .
\end{eqnarray}
In the limit $E \rightarrow 0$ one recovers the usual Planckian spectrum as first derived by Hawking. However, because of the non-trivial $E$ dependence of the action, the emission spectrum clearly deviates from that of a black body.

The emission spectrum follows from the tunneling  rate using the method presented in \cite{Vanzo:2011wq} and it deviates from that of a black body as can easily be seen from
\begin{equation}
    F(\omega)=\frac{d\omega}{2\pi}\frac{1}{e^{8 \pi  \omega M  G_N \left (1-\frac{\omega}{2 M}\right)
    \left (1-\frac{2\pi c_6}{G_N^2 M^2\left(M-\omega\right)^2}\right )}-1} 
\end{equation}
because of its additional dependence on $\omega$.
For small values of $\omega$, the above expression reduces to the Planck distribution with the modified Hawking temperature given by Eq. \eqref{PTm}. 

These results show that there is information about the interior state in the radiation from a black hole. Not only do the Hawking amplitudes depend on the quantum corrections to the metric (quantum hair), but the power spectrum also depends on these corrections and it does not match that of a black body once energy conservation and quantum gravitational effects  are taken into account.

\section{Arbitrary Quantum Hair and Hawking radiation}

It has been shown in \cite{Calmet:2022bpo} that quantum hair is a generic feature of quantum gravity. In particular, any collapse model will have quantum hair that carries information about the system throughout the gravitational collapse. The quantum hair captures the memory of the collapse process and of all the matter that formed the black hole. We can parametrize generic quantum hair as
\begin{eqnarray}\label{gencor}
ds^2 &=& -f(t,r) dt^2 +\frac{1}{g(t,r)} dr^2 + r^2 d\Omega^2 \, , \\ \nonumber
d\Omega^2&=& d\theta^2 +\sin(\theta)^2 d\phi^2 \, ,\\ \nonumber
f(t,r)&=& 1- \frac{ 2 G_N M}{r} + f^{(q)}(t,r) \, , \\ \nonumber
g(t,r)&=& 1- \frac{ 2 G_N M}{r} + g^{(q)}(t,r) \, , 
\end{eqnarray}
where $ f^{(q)}(r,t)$ and $ g^{(q)}(r,t)$ are assumed to be subleading corrections.
Note that this metric applies to any space-time dependent collapse model which admits the exterior Schwarzschild solution as a classical solution outside the matter distribution. We assume that the quantum correction can depend on time, but the collapse is expected to happen swiftly and the black hole thus evolves quickly to an equilibrium Schwarzschild-like configuration. Although we retain explicit space-time dependence in the quantum correction, we assume that the quantum correction is slowly varying over the Hawking emission timescale, so that the derivation based on geodesics is a reasonable approximation.

To leading order in the quantum correction, the horizon radius is given by
\begin{eqnarray}
 r_H=r_S\left(1- g^{(q)}(t,r_S)\right),
 \end{eqnarray}
 where $r_S=2 G_N M$. 

Starting from Eqs.~\eqref{eq:drdl} and~\eqref{eq:dudl} and taking $f(r)$ and $g(r)$ as in~\eqref{gencor}, we perform the integral of the square root of~\eqref{eq:drdl} in the interval $r'\in\left[r_H,r\right]$  corresponding to $\lambda'\in\left[0,\lambda\right]$. Using this procedure, we can obtain the coordinate $u(v)$  which depends on the quantum corrections $f^{(q)}(r,t)$ and $g^{(q)}(r,t)$.

It is thus obvious that the quantum corrected Bogoliubov coefficients depend on the quantum corrections as well as can be seen from the general expressions
\begin{eqnarray}
    \alpha_{\omega\omega'}=K
    \int_{-\infty}^{v_0} \,dv\,\Big(\frac{\omega'}{\omega}\Big)^{1/2}e^{i\omega^\prime v}e^{-i\omega u(v)}
\end{eqnarray}
and
\begin{eqnarray}
    \beta_{\omega\omega'}=K
    \int_{-\infty}^{v_0} \,dv\,\Big(\frac{\omega'}{\omega}\Big)^{1/2}e^{-i\omega^\prime v}e^{-i\omega u(v)}.
\end{eqnarray}
Thus the amplitude to emit a Hawking particle depends on the quantum correction to the metric, i.e., the quantum hair. Information about the collapsing matter is stored in the quantum correction to the gravitational field and it is then imprinted in Hawking radiation. Clearly generic quantum corrections would also leave an imprint in the modified emission spectrum obtained by imposing the energy conservation constraint as done by Parikh and Wilczek and applied to the case of a quantum corrected Schwarzschild metric in section~\ref{sec:3}.

\section{Conclusions}

In this paper we have considered quantum gravitational corrections to the emission of Hawking particles. We first describe a black hole using a Schwarzschild metric as originally done by Hawking. Extending his work to take into account known quantum gravitational corrections to the metric, we find that the quantum amplitude for the emission of a Hawking particle depends on the quantum hair, i.e., on the quantum gravitational correction to the classical Schwarzschild background. 

We then apply the method of Parikh and Wilczek who demonstrated that, when energy conservation is applied to the evaporation of a black hole, the thermal spectrum deviates from that of a black body. We calculate quantum gravitational corrections to the power spectrum and again demonstrate that the spectrum depends on the quantum hair.

Finally, we study a generic quantum gravitational correction which can parametrize the quantum corrections to any space-time dependent collapse model which admits the exterior Schwarzschild solution outside the matter distribution. We show that the quantum hair generically affects the quantum Hawking amplitude. Information about the collapsing matter is stored in the quantum correction to the gravitational field and it is then imprinted in Hawking radiation. 

In order to perform explicit calculations we considered quantum hair which manifests as a semiclassical correction to the exterior metric. In general, as discussed in \cite{Calmet:2021stu,Calmet:2021cip,Calmet:2022swf}, the quantum state of the black hole interior is reflected in the quantum state $\Psi_g$ of its exterior graviton field. $\Psi_g$ will affect radiation amplitudes: i.e., it represents the background or environment in which the Hawking radiation originates. This is the mechanism by which information escapes the black hole and the reason why its evaporation is unitary. 

\bigskip
{\it Acknowledgments}: The work of X.C. is supported in part  by the Science and Technology Facilities Council (grants numbers ST/T00102X/1 and ST/T006048/1).

{\it Data Availability Statement:}
This manuscript has no associated data. Data sharing not applicable to this article as no datasets were generated or analysed during the current study.

\end{document}